\documentclass[12pt]{article}

\newcommand{\R}{{\sf R\hspace*{-0.9ex}\rule{0.15ex}%
{1.5ex}\hspace*{0.9ex}}}

\begin{document}
\title{Stargenfunctions, generally parametrized systems and a causal formulation of phase space quantum mechanics}
\author{
Nuno Costa Dias\footnote{{\it ncdias@mail.telepac.pt}} \\ 
Jo\~ao Nuno Prata\footnote{{\it joao.prata@ulusofona.pt}} \\ 
{{\it Departamento de Matem\'{a}tica}} \\
{{\it Universidade Lus\'ofona de Humanidades e Tecnologias}} \\ {{\it Av. Campo Grande, 376,  
1749-024 Lisboa, Portugal}}}
\date{}
\maketitle

\begin{abstract}
We address the deformation quantization of generally parametrized systems displaying a natural time variable. The purpose of this exercise is twofold: first, to illustrate through a pedagogical example the potential of quantum phase space methods in the context of constrained systems and particularly of generally covariant systems. Second, to show that a causal representation for quantum phase space quasidistributions can be easily achieved through general parametrization. This result is succinctly discussed.
\end{abstract}
\section{Introduction.}

Generally covariant systems are a particular kind of dynamical systems in which time is included among the canonical variables \cite{Henneaux}. The state of the system evolves along orbits parametrized by an unphysical scalar parameter. The formalism is invariant under reparametrizations of this parameter (for this reason these systems are also called general parametrized systems) leading to a gauge symmetry, which in the Hamiltonian formulation is implemented by a first class (Hamiltonian) constraint. One of the most striking properties of generally covariant theories is that the Hamiltonian is identically zero on the constraint hypersurface. Consequently there is no standard Hamiltonian time evolution (which coincides exactly with the gauge transformation). Instead, dynamics is to be found among the relations between the canonical variables that are determined by the Hamiltonian constraint \cite{Rovelli1,Gambini1,Rovelli2}. In other words, dynamics is the unfolding of the gauge transformation. The most famous example of a generally covariant theory is, of course, general relativity \cite{Kuchar1,Ashtekar1,Rovelli3}.

Upon quantization these systems display a number of technical and conceptual problems which, motivated by the quest to quantize general relativity, have been intensively studied in the literature. Since the publication of Dirac's seminal work on the quantization of constrained systems \cite{Dirac}, several quantization programs have been developed with the aim of refining Dirac's approach and making it suitable to address the generally covariant case \cite{Henneaux,Ashtekar1,Ashtekar2,Marolf,Woodhouse}. In spite of this, several important problems concerning the quantization of these systems (the problem of time, the observables' problem and the measurement problem, just to mention a few) are still lacking a definitive answer \cite{Gambini1,Rovelli2,Marolf}.

One quantization program that has only been scarcely explored in this context is the deformation approach \cite{Weyl}-\cite{Bayen2}. And this regardless of the fact that the deformation quantization method displays some remarkable features that make it especially suitable to address the generally covariant case \cite{Antonsen,Martinez,Takao}.    
On the one hand the deformation approach leads to a formulation of quantum mechanics in terms of classical like objects. The theory lives in phase space and mimics the structure of classical statistical mechanics. The state of the system is described by a phase space quasi-distribution and observables are also functionals on the phase space. From the deformation point of view quantization amounts to the substitution of the standard product of functions by a non-commutative star product \cite{Bayen1,Bayen2}. The theory has been proved useful when addressing a wide range of fields of research ranging from topics in non-relativistic quantum mechanics \cite{Lee1}, \cite{Lee2}-\cite{Shin} to some current developments in $M$-theory \cite{Fairlie2,Fairlie3,Seiberg}. 
In the context of generally covariant systems it has been advocated that the relation between classical statistical and standard operator quantum mechanics  should be emphasized when approaching the quantization of these systems \cite{Kuchar2}. The point of view is that the problems besetting their quantization, being to a large extend conceptual ones, may receive key physical insights from the more intuitive classical statistical analysis. From this perspective the deformation methods seem to be especially suited.

On the technical side the deformation approach acquired the status of a powerful mathematical theory not displaying any of the subtleties of the path integral or canonical quantization prescriptions. Most remarkable is the fact that the theory is able to generate the quantum version of a classical system living on a generic (possibly curved) Poisson or sympletic manifold \cite{Kontsevich,Fedosov,Batalin,Bordemann}.    
Of special relevance for generally covariant systems is that the deformation approach displays some powerful tools to address the quantum eigenvalue problem, the most meaningful of which being a covariant formalism unifying all distinct phase space representations of the eigenvalue problem \cite{Bayen1,Nuno6,Nuno7}, a formal solution of a generic stargenvalue equation \cite{Nuno1,Dahl} and a set of efficient methods to determine the semiclassical expansion of the stargenfunctions \cite{Almeida1,Almeida2}.

In this paper we aim at providing a pedagogical incursion into some of these methods by studying the deformation quantization of a special simple kind of generally covariant systems: those obtained by parametrizing an originally non-covariant version of the system. The parametrization procedure leads to a new formulation of the dynamics living on an enlarged phase space where the physical time is incorporated as a canonical variable \cite{Henneaux}. 
The existence of such a natural time variable considerably simplifies the (technical and conceptual) analysis of the dynamical structure of the system. For this reason these models have been used \cite{Henneaux,Gambini1,Rovelli2} to test a set of quantization techniques and interpretative prescriptions aiming at addressing the far more difficult problem of quantizing "already parametrized systems" of which the most significant example is general relativity. 

The Hamiltonians of these models (as in the general case) are identically zero on the constraint hypersurface. At the quantum level the imposition of this constraint determines the physical states of the system. 
 In the deformation context the imposition of first class constraints yields a stargenvalue equation \cite{Nuno1,Fairlie1} and so the physical states are the zero stargenfunctions of the Hamiltonian symbol. We are then able to apply the powerful tools of the deformation approach to study the quantum eigenvalue problem. A complete characterization of the physical states will be provided in three different phase space representations. In particular, we obtain a representation where the dynamics of the quasidistribution (with respect to the physical time variable) is dictated by the classical Liouville equation. This is an interesting side result that shows that through general parametrization a causal formulation for the distributional sector of quantum mechanics is made possible.   
More generally we find that the remarkable parallelism between classical statistical mechanics and phase space quantum mechanics carries on intact to the generally covariant setting, a property that establishes a promising starting point for the future research on the deformation quantization of generally covariant systems.

\section{Covariant Wigner quantum mechanics}

The aim of this section is to review the main structures of the covariant formulation of the deformation quantization procedure. Our analysis will be restricted to the case of flat phase space as we will be dealing only with systems of this sort. The reader should refer to \cite{Batalin,Nuno6} for a more detailed presentation of these results and to \cite{Fedosov,Bordemann} for the generalization of the formalism to the non-flat case. An important part of this section (eq.(9) to eq.(15)) focuses on the covariant generalization of the $*$-genvalue equation, once again, just for the flat case. This is a topic that will play an important role latter on.  

Before proceeding and to avoid future misunderstandings let us make the following remark: Throughout the paper the word {\it covariant} will be used in two different contexts: to designate the invariance under time reparametrizations of the original classical system and to designate the invariance under general coordinate transformations of the Moyal plane. The aim of this paper is to present the deformation quantization of a generally covariant (parametrized) classical system. We use the covariant version of the deformation quantization procedure (a subject where this paper makes no original contribution) because this leads to a larger set of possible quantum phase space representations, among which we will find the causal representation presented in section 4. 

Let us then settle down the preliminaries: we consider a $N$-dimensional dynamical system, its classical formulation living on the flat phase space $T^*M$. A set of global canonical coordinates ($\{q_i,p_i,i=1..N\}$) can then be defined on $T^*M$ in terms of which the sympletic structure reads $w=dq_i \wedge dp_i$. 
Upon quantization the set $\{\hat q_i \} $ constitutes a complete set of commuting observables. Let us introduce the vector notation $\hat{\vec q}=(\hat q_1,...,\hat q_N)$ and designate by $|\vec q>$ the general eigenstate of $\hat{\vec q}$ associated to the array of eigenvalues $\vec q$ and spanning the Hilbert space ${\cal H}$ of the system. Let also ${\cal A}({\cal H})$ be the algebra of quantum observables over ${\cal H}$ and ${\cal A}(T^*M)$ the algebra of classical functions over the classical phase space $T^*M$.  Let $(\vec Q,\vec P)$ be a second set of phase space coordinates, related to $(\vec q,\vec p)$ by a generic phase space diffeomorphism (not necessarily canonical): $\vec q=\vec q(\vec Q,\vec P)$ and $\vec p=\vec p(\vec Q,\vec P)$. The covariant Weyl transform \cite{Nuno6}: 
\begin{eqnarray}
W^{(\vec q,\vec p)}_{(\vec Q,\vec P)}: {\cal A}({\cal H}) & \longrightarrow & {\cal A}(T^*M);  \\
\hat A & \longrightarrow & W^{(\vec q,\vec p)}_{(\vec Q,\vec P)}(\hat A) = \hbar^N \int d^N\vec x \int d^N\vec y \, e^{-i\vec p(\vec Q,\vec P) \cdot \vec y}
\delta (\vec x-\vec q(\vec Q,\vec P))
<\vec x+\frac{\hbar}{2} \vec y| \hat A |\vec x-\frac{\hbar}{2} \vec y> \nonumber
\end{eqnarray}
where $|\vec x \pm \frac{\hbar}{2} \vec y>$ are eigenstates of $\hat{\vec q}$, 
yields the entire structure of covariant phase space quantum mechanics. 
The map $W^{(\vec q,\vec p)}_{(\vec Q,\vec P)}$ can be applied both to an observable $\hat A$ as well as to the density matrix $|\psi(t)><\psi(t)|$. In the first case it yields the $^{(\vec q,\vec p)}_{(\vec Q,\vec P)}$-Weyl symbol $A'(\vec Q,\vec P)=W^{(\vec q,\vec p)}_{(\vec Q,\vec P)}(\hat A)$ of the original quantum operator and, in the second case, the celebrated Wigner function $f'_W(\vec Q,\vec P,t)=\frac{1}{(2\pi \hbar)^N}W^{(\vec q,\vec p)}_{(\vec Q,\vec P)}(|\psi(t)><\psi(t)|)$. Notice that if $(\vec Q,\vec P)=(\vec q,\vec p)$ then the covariant map $W^{(\vec q,\vec p)}_{(\vec Q,\vec P)}$ reduces to the standard Weyl transform in the variables $(\vec q,\vec p)$. In this case we will use the notation $W_{(\vec q,\vec p)}$ to designate $W_{(\vec q,\vec p)}^{(\vec q,\vec p)}$. Notice also that $A'(\vec Q,\vec P)=A(\vec q(\vec Q,\vec P),\vec p(\vec Q,\vec P))$ where $A(\vec q,\vec p)=W_{(\vec q,\vec p)}(\hat A)$ and the same relation is valid for the Wigner function.  
The covariant Weyl map implements the transformation $(\vec q,\vec p) \to (\vec Q,\vec P)$ as a coordinate transformation in quantum phase space and defines a covariant star-product $*'_{(\vec Q,\vec P)}$ and Moyal bracket $[,]_{M'_{(\vec Q,\vec P)}}$ \cite{Bayen1,Nuno6} ($\hat A,\hat B \in {\cal A}({\cal H})$):
\begin{eqnarray}
W^{(\vec q,\vec p)}_{(\vec Q,\vec P)}(\hat A\hat B) & = & W^{(\vec q,\vec p)}_{(\vec Q,\vec P)}(\hat A)*'_{(\vec Q,\vec P)} W^{(\vec q,\vec p)}_{(\vec Q,\vec P)}(\hat B)\nonumber \\
W^{(\vec q,\vec p)}_{(\vec Q,\vec P)}(\frac{1}{i\hbar}[\hat A,\hat B]) &=& [W^{(\vec q,\vec p)}_{(\vec Q,\vec P)}(\hat A),W^{(\vec q,\vec p)}_{(\vec Q,\vec P)}(\hat B)]_{M'_{(\vec Q,\vec P)}}
\end{eqnarray}
which display the functional form:
\begin{eqnarray}
A'( \vec Q, \vec P) *'_{(\vec Q,\vec P)} B'( \vec Q, \vec P)& = & A'( \vec Q, \vec P) e^{\frac{i \hbar}{2} {\buildrel { \leftarrow}\over\nabla'}_i  J'^{ij}_{(\vec Q,\vec P)}  {\buildrel
{ \rightarrow}\over\nabla'}_j} B' (\vec Q, \vec P) \nonumber \\
\left[ A'(\vec Q,\vec P), B'(\vec Q,\vec P) \right]_{M'_{(\vec Q,\vec P)}} & = & \frac{2}{\hbar} A'(\vec Q,\vec P) \sin \left(\frac{\hbar}{2}
{\buildrel { \leftarrow}\over\nabla'}_i  J'^{ij}_{(\vec Q,\vec P)}  {\buildrel
{ \rightarrow}\over\nabla'}_j \right) B'(\vec Q,\vec P),
\end{eqnarray}
where the covariant derivative is given by (let $O'^i = P_i, O^i=p_i, i=1, \cdots ,N$; $O'^i = Q_{i-N}, O^i=q_{i-N}, i=N+1, \cdots, 2N$):
\begin{equation}
\nabla'_i A'  = \partial'_i A', \quad
\nabla'_i \nabla'_j A'  = \partial'_i \partial'_j A' - \Gamma'^k_{ij} \partial'_k A', \quad \partial'_i= \partial /\partial {O'}^{i};
\quad i,j,k= 1, \cdots, 2N,
\end{equation}
and
\begin{equation}
J'^{ij}_{(\vec Q,\vec P)}(\vec Q,\vec P) = \frac{\partial O'^i}{\partial O^k} \frac{\partial O'^j}{\partial O^l} J^{kl}_{(\vec q,\vec p)}, \quad 
\Gamma'^i_{jk}(\vec Q,\vec P) =  \frac{\partial O'^i}{\partial O^b} \frac{\partial^2 O^b}{\partial O'^j \partial O'^k},
\end{equation}
are the new symplectic structure and Poisson connection associated with the coordinates $(\vec Q,\vec P)$, respectively. Finally, notice that in eq.(5) we explicitly took into account the phase space flat structure.

When formulated in terms of these structures Wigner mechanics becomes fully invariant under the action of general phase space diffeomorphisms:
\begin{equation}
A'(\vec Q,\vec P) *^{\prime}_{(\vec Q,\vec P)} B'(\vec Q,\vec P) = A(\vec q(\vec Q,\vec P),\vec p(\vec Q,\vec P))*_{(\vec q,\vec p)} B(\vec q(\vec Q,\vec P),\vec p(\vec Q,\vec P)) \qquad \forall_{A,B \in {\cal A}(T^{\ast}M)},
\end{equation}
the covariant generalization of the Moyal and stargenvalue equations reading:
\begin{eqnarray}
& \frac{\partial f'_W}{\partial t}  = [H',f'_W]_{M'_{(\vec Q,\vec P)}} & \nonumber \\
& \left\{ \begin{array}{lll}
A'(\vec Q,\vec P) *'_{(\vec Q,\vec P)} \rho'_{a;b}(\vec Q,\vec P) & = &  a \rho'_{a;b}(\vec Q,\vec P) \\
\rho'_{a;b}(\vec Q,\vec P) *'_{(\vec Q,\vec P)} A'(\vec Q,\vec P) & = &  b \rho'_{a;b}(\vec Q,\vec P)
\end{array} \right. &
\end{eqnarray}
where $\rho'_{a;b}(\vec Q,\vec P)$ is the $a$-left and $b$-right $*'_{(\vec Q,\vec P)}$-genfunction of $A'(\vec Q,\vec P)$.
These equations transform covariantly under arbitrary phase space diffeomorphisms yielding, in any coordinates, identical mathematical solutions and thus identical physical predictions:
\begin{equation}
P(A'(\vec Q,\vec P;t)=a)=  \int d^N\vec Q \int d^N\vec P (\mbox{det} J'^{ij}_{(\vec Q,\vec P)})^{-1/2}
\rho'_{a;a}(\vec Q,\vec P)  f'_W(\vec Q,\vec P;t).
\end{equation}

Let us consider the stargenvalue equation (7) in more detail. Let $\{\hat A_1,...,\hat A_N\}$ and $\{\hat B_1,...,\hat B_N\}$ be two sets of commuting observables satisfying $[\hat A_i,\hat B_j]=i \hbar \delta_{ij}$. Let $A'_i(\vec Q,\vec P)=W^{(\vec q,\vec p)}_{(\vec Q,\vec P)}(\hat A_i)$ and $B'_i(\vec Q,\vec P)=W^{(\vec q,\vec p)}_{(\vec Q,\vec P)}(\hat B_i)$. Then the simultaneous solution of:
\begin{equation}
\left\{ \begin{array}{lll}
A'_i(\vec Q,\vec P) *'_{(\vec Q,\vec P)} \rho'_{a_1,..,a_N;b_1,..,b_N}(\vec Q,\vec P)& = & a_i \rho_{a_1,..,a_N;b_1,..,b_N}(\vec Q,\vec P) \\
\rho'_{a_1,..,a_N;b_1,..,b_N}(\vec Q,\vec P) *'_{(\vec Q,\vec P)} A'_i(\vec Q,\vec P)
& = & b_i \rho_{a_1,..,a_N;b_1,..,b_N}(\vec Q,\vec P) \quad , \quad i=1..N
\end{array} \right.
\end{equation}
is given by \cite{Nuno1}:
\begin{eqnarray}
\rho'_{a_1,..,a_N;b_1,..,b_N}(\vec Q,\vec P) & \equiv & \Delta_{*'_{(\vec Q,\vec P)}}(A'_1,...,A'_N;a_1,...,a_N;b_1,...,b_N)   ={\prod_{i=1}^{N}}{}_{*'_{(\vec Q,\vec P)}}  \rho'_{a_i;b_i}(\vec Q,\vec P), \\
\rho'_{a_i;b_i}(\vec Q,\vec P) & \equiv & \Delta_{*'_{(\vec Q,\vec P)}}(A'_i;a_i;b_i) = \frac{1}{2\pi}
e_{*'_{(\vec Q,\vec P)}}^{\frac{i}{\hbar}(b_i-a_i)B'_i(\vec Q,\vec P)}*'_{(\vec Q,\vec P)}
\int \,dk e_{*'_{(\vec Q,\vec P)}}^{ik(A'_i(\vec Q,\vec P)-b_i)},
\end{eqnarray}
where ${\prod_{i=1}^{N}}{}_{*'_{(\vec Q,\vec P)}} $ is the $N$-fold starproduct and the starexponential $e_{*'_{(\vec Q,\vec P)}}^{ikA'_i(\vec Q,\vec P)} \equiv E_{*'_{(\vec Q,\vec P)}}(k,\vec Q,\vec P)$ is defined as the solution of the differential problem:
$ \frac{\partial}{\partial k} E_{*'_{(\vec Q,\vec P)}}(k,\vec Q,\vec P)=i A_i'(\vec Q,\vec P) *'_{(\vec Q,\vec P)} E_{*'_{(\vec Q,\vec P)}}(k,\vec Q,\vec P)$ and $ E_{*'_{(\vec Q,\vec P)}}(0,\vec Q,\vec P)=1$, which is unique  for the Weyl symbol $A'_i(\vec Q,\vec P)$ of a generic self-adjoint operator $\hat A_i$ (a result that is just the Weyl-Wigner translation of an equivalent result in operator quantum mechanics \cite{Reed}). Furthermore $ E_{*'_{(\vec Q,\vec P)}}(k,\vec Q,\vec P)= \sum_{n=1}^{+\infty} \frac{(ik)^n}{n!} {\prod_{j=1}^{N}}{}_{*'_{(\vec Q,\vec P)}} A'_i(\vec Q,\vec P)$ whenever $k$ is in the radius of convergence of the series, which justifies the notation used in eq.(11). We also introduced the $*$-delta notation
$\Delta_{*'_{(\vec Q,\vec P)}}(A'_1,...,A'_N;a_1,...,a_N;b_1,...,b_N) $ that will be extensively used later on. Notice that the stargenfunctions $\rho'_{a_1,..,a_N;b_1,..,b_N}(\vec Q,\vec P)$ are the $_{(\vec Q,\vec P)}^{(\vec q,\vec p)}$-Weyl transform of the projectors $|a_1,..,a_N><b_1,..,b_N|$ where $|a_1,..,a_N>$ and $|b_1,..,b_N>$ are simultaneous eigenvectors of $\hat A_i$, $i=1..N$. Moreover the stargenfunctions $\rho'_{a_i;b_i}(\vec Q,\vec P)$ are the $_{(\vec Q,\vec P)}^{(\vec q,\vec p)}$-Weyl transform of the projector:
\begin{equation}
|a_i><b_i|=\int\, d\vec z\, |a_i, \vec z><b_i,\vec z|= \frac{1}{2\pi} \int \, dk e^{\frac{i}{\hbar}(b_i-a_i)\hat B_i} e^{ik(\hat A_i-b_i)},
\end{equation}
where $\vec z=(a_1,...,a_{i-1},a_{i+1},...,a_N)$ is the $N-1$ array of degeneracy indices. The Weyl transform of the first identity in eq.(12) yields the inverse relation of eq.(10):
\begin{equation} 
\rho'_{a_i;b_i}(\vec Q,\vec P)= \int \, da_1..da_{i-1}da_{i+1}..da_N \rho'_{a_1,..,a_i,..,a_N;a_1,..,b_i,..,a_N}(\vec Q,\vec P).
\end{equation}
Finally, the diagonal elements $\rho'_{a_i;a_i}(\vec Q,\vec P)$ are given by:
\begin{equation}
\rho'_{a_i;a_i}(\vec Q,\vec P)= \frac{1}{2\pi} \int \, dk e_{*'_{(\vec Q,\vec P)}}^{ik (A'_i(\vec Q,\vec P)-a_i)}=
\Delta_{*'_{(\vec Q,\vec P)}}(A'_i;a_i;a_i) \equiv  \Delta_{*'_{(\vec Q,\vec P)}}(A'_i-a_i),
\end{equation}
this being the object that enters the probability functional (8). 
The stargenfunctions (10,11) transform as scalars under arbitrary phase space coordinate transformations. For instance $
\rho'_{a_i;a_i}(\vec Q,\vec P)$ satisfies:
\begin{equation}
\rho'_{a_i;a_i}(\vec Q,\vec P)=\Delta_{*'_{(\vec Q,\vec P)}}(A'_i(\vec Q,\vec P)-a_i)=\Delta_{*_{(\vec q,\vec p)}}(A_i(\vec q(\vec Q,\vec P),\vec p(\vec Q,\vec P))-a_i)=\rho_{a_i;a_i}(\vec q(\vec Q,\vec P),\vec p(\vec Q,\vec P))
\end{equation}
where $\rho_{a_i;a_i}(\vec q,\vec p)$ is the diagonal solution of the stargenvalue equation in the $(\vec q,\vec p)$-representation.

Lastly, we should point out that the former results (in the presented form) are valid whenever we are able to provide the complete sets of commuting observables $\{\hat A_i\}$ and $\{\hat B_i\}$. In fact this restriction can be considerably weakened. It is crucial to the overall approach that the complete set $\{\hat A_i\}$ exists, but the requirement on the existence of the set $\{\hat B_i\}$ can be discarded, while preserving the validity of eqs.(8,10,13,14) exactly and that of eqs.(11,12) under slight modifications. This is the case, for instance, when $\hat A_i$ displays a discrete spectrum. The reader should refer to \cite{Nuno1} for a detailed presentation of these results.

\section{Deformation quantization of a parametrized non-relativistic system}

Let us consider an arbitrary $N$-dimensional dynamical system with configuration variables $\vec q=(q_1,...,q_N)$, described by the Lagrangian $L_0(\vec q,\frac{d\vec q}{dt})$.  Starting from the standard action $S=\int dt L_0(\vec q,\frac{d\vec q}{dt})$ we impose the time reparametrization invariance by introducing an unphysical time $\tau$ and promoting $t$ to a configuration variable. The action is now re-written as \cite{Henneaux}:
\begin{equation}
S=\int d \tau \, {\dot t} L_0 \left( \vec q,\frac{\dot{\vec q}}{\dot t} \right)
\end{equation}
where the dot represents the derivative with respect to $\tau$. The Legendre transform yields the Hamiltonian formulation living in a $(2N+2)$-dimensional phase space spanned by the canonical variables $t,P_t,\vec q,\vec p$ where $P_t=\partial L/ \partial \dot{t}$, $p_i=\partial L/ \partial \dot{q}_i$, $i=1..N$ and $L(t,\dot t,\vec q,\dot{\vec q})={\dot t} L_0 \left( \vec q,\frac{\dot{\vec q}}{\dot t} \right)$. This, as expected, is a "zero-Hamiltonian" system:
\begin{equation}
H=\lambda \phi \quad , \quad \phi = P_t+ H_0(\vec q,\vec p)
\end{equation}
where $\phi$ is a first class constraint, $\lambda $ is a Lagrange multiplier and $H_0$ is the Hamiltonian associated to the original Lagrangian $L_0$.  

Dirac's quantization procedure imposes the constraint as a restriction on the space of physical states: $ \hat{\phi} |\psi> =0$. In the density matrix formulation this equation reads: 
\begin{equation}
\hat{\phi} |\psi><\psi| = |\psi><\psi| \hat{\phi}= 0. 
\end{equation}
To find the complete specification of its solutions, and according to the general method of section 2, the first step is to introduce the complete set of observables $\{\hat{\phi},\hat{\vec A}=(\hat A_1,...,\hat A_N) \}$ and the set of operators $\hat{\vec B}=(\hat B_1,...,\hat B_N)$, generators of  translations in the spectrum of  $\hat{\vec A}$, satisfying $[\hat A_j,\hat B_k]=i\hbar \delta_{jk} $ and $[\hat{\phi},\hat B_k ]=0$, $j,k=1..N$. We easily find that:
\begin{equation}
\hat A_j=\hat F_j(\hat{\vec q},\hat{\vec p},-\hat t) \quad , \quad \hat B_k=\hat G_k(\hat{\vec q},\hat{\vec p},-\hat t)
\end{equation}
where $\hat F_j(\hat{\vec q},\hat{\vec p},t)$ and $\hat G_k(\hat{\vec q},\hat{\vec p},t)$ are the solutions of the original Heisenberg equations:
\begin{equation}
\frac{\partial \hat F_j}{\partial t}=[\hat F_j,\hat H_0] \quad , \quad  \frac{\partial \hat G_k}{\partial t}=[\hat G_k,\hat H_0] \quad, \quad j,k=1..N
\end{equation}
together with the initial conditions $\hat F_j(\hat{\vec q},\hat{\vec p},0)=\hat q_j$ and $\hat G_k(\hat{\vec q},\hat{\vec p},0)=\hat p_k$, satisfy the aforementioned requirements. $\hat{\vec A}$ and $\hat{\vec B}$ are the quantum histories of the system.

The phase space representation of eq.(18) is determined by the Weyl transform. We may write:
\begin{equation}
\phi *_{(t,\phi,{\vec A},{\vec B})} f_W (t,\phi,{\vec A},{\vec B}) = f_W(t,\phi,{\vec A},{\vec B}) *_{(t,\phi,{\vec A},{\vec B})} \phi =0
\end{equation}
by using the map $W_{(t,\phi,{\vec A},{\vec B})}$ or equivalently:
\begin{equation}
\phi(t,P_t,{\vec q},{\vec p}) *_{(t,P_t,{\vec q},{\vec p})} f_W (t,P_t,{\vec q},{\vec p}) = f_W(t,P_t,{\vec q},{\vec p}) *_{(t,P_t,{\vec q},{\vec p})} \phi(t,P_t,{\vec q},{\vec p}) =0
\end{equation}
by using the map $W_{(t,P_t,{\vec q},{\vec p})}$. We start by considering the simplest representation, which is provided by eq.(21). In this case the Wigner function is a left and right zero $(t,\phi,{\vec A},{\vec B})$-stargenfunction of the constraint symbol $\phi$. To determine $f_W$ explicitly we will follow the procedure described in section 2. From eqs.(10,11,14), the fundamental zero stargenfunctions of the Hamiltonian constraint (which will be designated by $\rho_{{\vec a},{\vec b}} \def \rho_{h=0,{\vec a};h=0,{\vec b}}$) are given by:
\begin{eqnarray}
\rho_{{\vec a},{\vec b}}(t,\phi,{\vec A},{\vec B}) & = & \Delta_{*_{(t,\phi,{\vec A},{\vec B})}}(\phi,{\vec A};h=0,{\vec a};h=0,{\vec b}) = \Delta_{*_{(t,\phi,{\vec A},{\vec B})}}(\phi) *_{(t,\phi,{\vec A},{\vec B})} \Delta_{*_{(t,\phi,{\vec A},{\vec B})}}({\vec A};{\vec a};{\vec b}) =\nonumber \\
& = & \Delta _{*_{(t,\phi,{\vec A},{\vec B})}}(\phi ){\prod_{j=1}^{N}}{}_{
{*_{(t,\phi,{\vec A},{\vec B})}}}\left\{ e_{*_{(t,\phi,{\vec A},{\vec B})}}^{\frac{i}{\hbar} (b_j-a_j)B_j} *_{(t,\phi,{\vec A},{\vec B})} \Delta _{*_{(t,\phi,{\vec A},{\vec B})}}(A_j-b_j) \right\}\nonumber \\
& = & \delta(\phi) {*_{(t,\phi,{\vec A},{\vec B})}}\prod_{j=1}^{N} \left\{ e^{\frac{i}{\hbar} (b_j-a_j)B_j} *_{(t,\phi,{\vec A},{\vec B})} \delta(A_j-b_j) \right\}\nonumber \\
& = & \delta (\phi) \prod_{j=1}^N \left\{ \frac{1}{2\pi} \sum_{n=0}^{+\infty} \frac{1}{n!} \left( \frac{-i\hbar}{2} \right)^n e^{\frac{i}{\hbar} (b_j-a_j)B_j} \left[\frac{i}{\hbar} (b_j-a_j) \frac{\partial }{\partial A_j} \right]^n \int dk \, e^{ik(A_j-b_j)} \right\} \nonumber \\
& = & \frac{\delta(\phi)}{(2\pi)^N}\prod_{j=1}^N \left\{ e^{\frac{i}{\hbar} (b_j-a_j)B_j} \int \, dk \sum_{n=0}^{+\infty} \frac{1}{n!} \left[\frac{i}{2}k(b_j-a_j)\right]^n e^{ik(A_j-b_j)} \right\} \nonumber \\
& = & \delta(\phi) \prod_{j=1}^N \left\{ e^{\frac{i}{\hbar} (b_j-a_j)B_j} \frac{1}{2\pi} \int \, dk  e^{ik(A_j-b_j+\frac{b_j}{2}-\frac{a_j}{2})}\right\} \nonumber \\
& = & \delta(\phi) \prod_{j=1}^N \left\{ e^{\frac{i}{\hbar} (b_j-a_j)B_j} \delta(A_j-\frac{a_j}{2}-\frac{b_j}{2}) \right\} 
 =  \delta(\phi) e^{\frac{i}{\hbar} (\vec b-\vec a)\cdot \vec B} \delta(\vec A-\frac{\vec a+\vec b}{2})
\end{eqnarray}
where ${\prod_{j=1}^{N}}{}_{
{*_{(t,\phi,{\vec A},{\vec B})}}}$ stands for the $N$-fold starproduct and $\prod_{j=1}^{N}$ for the standard $N$-fold product of functions. The stargenfunctions $\rho_{{\vec a},{\vec b}}(t,\phi,{\vec A},{\vec B})$ are simultaneously solutions of eq.(21) and:
\begin{equation}
\left\{ \begin{array}{lll}
A_j*_{(t,\phi,{\vec A},{\vec B})}\rho_{{\vec a},{\vec b}}(t,\phi,{\vec A},{\vec B}) & = & a_j \rho_{{\vec a},{\vec b}}(t,\phi,{\vec A},{\vec B}) \\
\rho_{{\vec a},{\vec b}}(t,\phi,{\vec A},{\vec B}) *_{(t,\phi,{\vec A},{\vec B})} A_j & = & b_j \rho_{{\vec a},{\vec b}}(t,\phi,{\vec A},{\vec B}) \quad , j=1..N
\end{array} \right.
\end{equation}
Notice that the diagonal elements ${\vec a}={\vec b}$ fully identify a history of the system: $\rho_{{\vec a},{\vec a}}(t,\phi,{\vec A},{\vec B})=\delta(\phi)\prod_{j=1}^N \delta(A_j-a_j)$. 
It is also straightforward to realize that the fundamental stargenfunctions (23) are the $(t,\phi,{\vec A},{\vec B})$-Weyl transform of the projectors:
\begin{equation}
|h=0,{\vec a}><h=0,{\vec b}| = \hat{\Delta}(\hat{\phi})\prod_{j=1}^N\left\{ e^{\frac{i}{\hbar} (b_j-a_j) \hat B_j}  \hat{\Delta} (\hat A_j-b_j) \right\}
\end{equation}
where the general ket satisfies:
\begin{equation}
\hat{\phi} |h,{\vec x}> =h|h,{\vec x}> \quad \mbox{and} \quad \hat A_j |h,{\vec x}> = x_j |h,{\vec x}>
\end{equation}
and 
\begin{equation}
\hat{\Delta}(\hat{\phi})= \int d{\vec x} |h=0,{\vec x}><h=0,{\vec x}| = \frac{1}{2\pi} \int dk e^{ik \hat{\phi}},
\end{equation}
is the operator analogue of (14).
The most general solution of eq.(21) is a linear combination of the fundamental solutions (23):
\begin{equation}
f_W(t,\phi,{\vec A},{\vec B}) =\frac{1}{(2\pi \hbar)^N}  \int d{\vec a} \int d{\vec b} C({\vec a}) C^* ({\vec b}) \rho_{{\vec a},{\vec b}} (t,\phi,{\vec A},{\vec B})
\end{equation}
where $C({\vec a})$ obeys to the normalization condition that is induced by the normalization of the Wigner function. Let us then calculate its norm:
\begin{eqnarray}
&& \int \int \int \int dtd\phi d{\vec A}d{\vec B} f_W (t,\phi,{\vec A},{\vec B})  
\nonumber \\
&=&  \frac{1}{(2\pi \hbar)^N} \int \int \int dt d{\vec A}d{\vec B} \int d{\vec a} \int d{\vec b} 
C({\vec a}) C^* ({\vec b})\prod_{j=1}^N e^{\frac{i}{\hbar} (b_j-a_j)B_j}  \delta (A_j-\frac{a_j+b_j}{2}) = \nonumber \\
&=& \int dt \int d{\vec a} \int d{\vec b} C({\vec a})C^*({\vec b})  \delta (\vec b-\vec a) = 
\int dt \int d{\vec a} C({\vec a})C^*({\vec a})  
\end{eqnarray}
The divergence of the previous integral indicates that we are integrating over the gauge orbits thus spoiling the normalization of the Wigner function. In this case this is quite simple to correct. In fact we just have to introduce the phase space measure $d\mu = d\phi d{\vec A}d{\vec B}$ and use it from now on whenever we have to integrate the Wigner function. The procedure corresponds to cutting the gauge orbits through a single time hypersurface. With this measure the Wigner function satisfies:
\begin{equation}
\int d \mu f_W = 1 \quad, \forall t
\end{equation}
provided the parameters $C({\vec a})$ satisfy $\int d{\vec a} | C({\vec a})|^2 =1$. The proper normalization of the Wigner function determines a new phase space measure and a restriction on the factors $C({\vec a})$. These parameters display a natural physical interpretation. To see this explicitly let us calculate the probabilities for the output of a measurement of ${\vec A}$. The general ${\vec x}=(x_1,...,x_N)$-left and -right stargenfunction of ${\vec A}$ (and simultaneously $y$-left and -right stargenfunction of the constraint) is:
\begin{eqnarray}
 \rho_{y,{\vec x};y,{\vec x}}  & = & \Delta_{*_{(t,\phi,{\vec A},{\vec B})}}(\phi,{\vec A};y,{\vec x};y,{\vec x}) = \Delta_{*_{(t,\phi,{\vec A},{\vec B})}}(\phi -y) {\prod_{j=1}^N}{}_{*_{(t,\phi,{\vec A},{\vec B})}} \Delta_{*_{(t,\phi,{\vec A},{\vec B})}} (A_j-x_j)=\nonumber \\  
& = & \delta (\phi -y)\prod_{j=1}^N \delta (A_j-x_j) =\delta (\phi -y) \delta (\vec A-\vec x)
\end{eqnarray}
Therefore, the probability density for ${\vec A}={\vec x}$ and $\phi=y$ is given by:
\begin{eqnarray}
&& {\cal P}(\phi=y, {\vec A}= {\vec x} )  =  \int d \mu \, f_W(t,\phi,{\vec A},{\vec B}) \delta (\phi-y) \delta ({\vec A}-{\vec x}) = \int \, d{\vec B} f_W(t,y,{\vec x},{\vec B}) \nonumber \\
& = &   \frac{1}{(2\pi \hbar)^N} \int \, d{\vec B} \int \int d{\vec a} d{\vec b}  C({\vec a}) C^* ({\vec b}) \delta (y) \prod_{j=1}^N \left\{ e^{\frac{i}{\hbar} (b_j-a_j)B_j}  \delta (x_j-\frac{a_j+b_j}{2}) \right\}\nonumber \\
& = &  \int \int d{\vec a} d{\vec b} \, C({\vec a})C^*({\vec b})  \delta({\vec b}-{\vec a}) \delta ({\vec x}-\frac{{\vec a}+{\vec b}}{2}) \delta (y) = |C({\vec x})|^2 \delta (y)
\end{eqnarray}
from where it follows:
\begin{equation}
P(\phi =y , {\vec A}={\vec x} ) = \lim_{\epsilon \to 0^+} \int_{y-\epsilon}^{y + \epsilon} dy' {\cal P} (\phi = y', {\vec A}={\vec x} ) = |C(\vec x)|^2 \delta_{y,0}. 
\end{equation}
It is clear from the previous equation that the term $|C({\vec x})|^2$ represents the probability for the system to be found in the history ${\vec A}={\vec x}$. Under the measurement of ${\vec A}$ with output ${\vec x}$ the Wigner function will indeed collapse to the state:
\begin{equation}
\rho_{0,{\vec x};0,{\vec x} } = \delta (\phi ) \delta (\vec A-\vec x) 
\end{equation}
Let us also point out that in the $(t,\phi,{\vec A},{\vec B})$-representation the Wigner function is static both with respect to the external time $\tau$ and to the physical time $t$. In fact from eqs.(23,28) we see that $\frac{\partial }{\partial t} f_W=0$. 

The former results can now be easily translated to the $(t,P_t,{\vec q},{\vec p})$ representation where the intention is to solve eq.(22). The two representations are related by the unitary transformation:
\begin{eqnarray}
t(t,P_t,{\vec q},{\vec p}) & = & U^{-1}*_{(t,P_t,{\vec q},{\vec p})}t*_{(t,P_t,{\vec q},{\vec p})}U=t \nonumber \\
\phi (t,P_t,{\vec q},{\vec p}) & = & W_{(t,P_t,{\vec q},{\vec p})}(\hat{\phi}) =U^{-1}*_{(t,P_t,{\vec q},{\vec p})}P_t*_{(t,P_t,{\vec q},{\vec p})}U=P_t+H_0({\vec q},{\vec p}) \nonumber \\
A_j (t,{\vec q},{\vec p}) & = & W_{(t,P_t,{\vec q},{\vec p})}(\hat{A_j}) =U^{-1}*_{(t,P_t,{\vec q},{\vec p})}q_j*_{(t,P_t,{\vec q},{\vec p})}U \quad, \quad j=1..N \nonumber \\
 B_j (t,{\vec q},{\vec p})& = & W_{(t,P_t,{\vec q},{\vec p})}(\hat{B_j}) =U^{-1}*_{(t,P_t,{\vec q},{\vec p})}p_j*_{(t,P_t,{\vec q},{\vec p})}U \quad , \quad j=1..N
\end{eqnarray}
where $U=e_{*_{(t,P_t,{\vec q},{\vec p})}}^{\frac{i}{\hbar}H_0({\vec q},{\vec p})t}$.The fundamental stargenfunctions are:
\begin{eqnarray}
&& \rho_{{\vec a},{\vec b}}(t,P_t,{\vec q},{\vec p})   =  \Delta_{*_{(t,P_t,{\vec q},{\vec p})}}(\phi,{\vec A};h=0,{\vec a};h=0,{\vec b}) \\
& = & \Delta_{*_{(t,P_t,{\vec q},{\vec p})}}(P_t+H_0({\vec q},{\vec p})) {\prod_{j=1}^N}{}_{*_{{(t,P_t,{\vec q},{\vec p})}}}\left\{
e_{*_{(t,P_t,{\vec q},{\vec p})}}^{\frac{i}{\hbar} (b_j-a_j)B_j(t,{\vec q},{\vec p})} {*_{(t,P_t,{\vec q},{\vec p})}} \Delta _{*_{(t,P_t,{\vec q},{\vec p})}}(A_j(t,{\vec q},{\vec p})-b_j) \right\} \nonumber 
\end{eqnarray}
which in general do not simplify any further as in equation (23). These stargenfunctions are the ${(t,P_t,{\vec q},{\vec p})}$-Weyl transform of the projector (25). The most general solution of (22) is then:
\begin{equation}
f_W(t,P_t,{\vec q},{\vec p})=\frac{1}{(2\pi \hbar)^N} \int d{\vec a} d{\vec b} \, C({\vec a})C^*({\vec b}) \rho_{{\vec a},{\vec b}}(t,P_t,{\vec q},{\vec p})
\end{equation}
and can be obtained directly from the Wigner function (28) by applying the unitary transformation (35). Hence, $f_W(t,P_t,{\vec q},{\vec p})$ 
is properly normalized for the phase space measure $d\mu = dP_t d{\vec q}d{\vec p} $ provided the coefficients $C({\vec a})$ satisfy the normalization $\int d{\vec a} \, C({\vec a})C^*({\vec a})=1$. Notice that in general the new Wigner function does not have support exclusively on the classical constraint hypersurface. This is due to the non-local character of the $*$-delta functions in eq.(36). Notice also that $f_W$ does not 
evolve with respect to the external time $\tau$ but it displays the standard time evolution with respect to the canonical time $t$:
\begin{eqnarray}
\frac{\partial f_W(t,P_t,{\vec q},{\vec p})}{\partial \tau} & = & [H(t,P_t,{\vec q},{\vec p}),f_W(t,P_t,{\vec q},{\vec p})]_{M_{(t,P_t,{\vec q},{\vec p})}}=0 \nonumber \\
& \Longleftrightarrow & 
\frac{\partial f_W(t,P_t,{\vec q},{\vec p})}{\partial t}=[H_0({\vec q},{\vec p}),f_W(t,P_t,{\vec q},{\vec p})]_{M_{({\vec q},{\vec p})}} 
\end{eqnarray}
We conclude that the $(t,P_t,{\vec q},{\vec p})$-representation yields the extended phase space Schr\"odinger picture for the quantum generally covariant system. Likewise the $(t, \phi, {\vec A},{\vec B})$-representation provides the phase space Heisenberg picture.

\section{Causal representation}

We now study another possible phase space representation of the system. The first step is to specify a new set of classical phase space coordinates $(t,P_t,{\vec Q},{\vec P})$. Let us define the phase space diffeomorphism by:
\begin{equation}
t=t \quad ,\quad P_t=\phi-H_0({\vec A},{\vec B}) \quad , \quad {\vec Q}={\vec Q}(t,{\vec A},{\vec B}) \quad , \quad {\vec P}={\vec P}(t,{\vec A},{\vec B})
\end{equation}
where $H_0({\vec A},{\vec B})=W_{({\vec A},{\vec B})}(\hat H_0)=W_{({\vec q},{\vec p})}(\hat H_0)|_{{\vec q}={\vec A} \wedge {\vec p}={\vec B}}$ and ${\vec Q}(t,{\vec A},{\vec B})$, ${\vec P}(t,{\vec A},{\vec B})$ satisfy: 
\begin{equation}
\frac{\partial {\vec Q}(t,{\vec A},{\vec B})}{\partial t}=\{{\vec Q}(t,{\vec A},{\vec B}),H_0({\vec A},{\vec B})\}_{({\vec A},{\vec B})} \quad , \quad 
\frac{\partial {\vec P}(t,{\vec A},{\vec B})}{\partial t}=\{{\vec P}(t,{\vec A},{\vec B}),H_0({\vec A},{\vec B})\}_{({\vec A},{\vec B})}
\end{equation}
together with the initial conditions: ${\vec Q}(0,{\vec A},{\vec B})={\vec A}$ and ${\vec P}(0,{\vec A},{\vec B})={\vec B}$. That is ${\vec Q}(t,{\vec A},{\vec B})$ and ${\vec P}(t,{\vec A},{\vec B})$ constitute the {\it classical} time evolution of the deparametrized system. The transformation (39) is canonical and can be easily inverted. It yields:
\begin{equation}
t=t \quad ,\quad \phi=\phi'(t,P_t,{\vec Q},{\vec P})=P_t+H_0({\vec Q},{\vec P}) \quad , \quad {\vec A}={\vec A}'(t,{\vec Q},{\vec P}) \quad , \quad {\vec B}={\vec B}'(t,{\vec Q},{\vec P})
\end{equation}
where this time $H_0({\vec Q},{\vec P})=W_{({\vec q},{\vec p})}(\hat H_0)|_{{\vec q}={\vec Q} \wedge {\vec p}={\vec P}}$ and ${\vec A}'(t,{\vec Q},{\vec P})$, ${\vec B}'(t,{\vec Q},{\vec P})$ satisfy: 
\begin{equation}
\frac{\partial {\vec A}'(t,{\vec Q},{\vec P})}{\partial t}=\{H_0({\vec Q},{\vec P}),{\vec A}'(t,{\vec Q},{\vec P})\}_{({\vec Q},{\vec P})} \quad , \quad 
\frac{\partial {\vec B}'(t,{\vec Q},{\vec P})}{\partial t}=\{H_0({\vec Q},{\vec P}),B'(t,{\vec Q},{\vec P})\}_{({\vec Q},{\vec P})}
\end{equation}
together with the initial conditions: ${\vec A}'(0,{\vec Q},{\vec P})={\vec Q}$ and ${\vec B}'(0,{\vec Q},{\vec P})={\vec P}$. This means that the {\it functions} ${\vec A}'(t,{\vec Q},{\vec P})$ and ${\vec B}'(t,{\vec Q},{\vec P})$ are the classical histories of the system. We should point out that in general the two functions ${\vec A}'(t,{\vec Q},{\vec P})$ and ${\vec A}(t,{\vec q},{\vec p})$ given by eqs.(41,35), respectively, as well as ${\vec B}'(t,{\vec Q},{\vec P})$ and ${\vec B}(t,{\vec q},{\vec p})$, display different functional forms (they are respectively the classical and the quantum Weyl-Wigner histories of the system). To make this point explicit, we introduced the prime notation which is in agreement with the fact that ${\vec A}'(t,{\vec Q},{\vec P})=W_{(t,P_t,{\vec Q},{\vec P})}^{(t,\phi,{\vec A},{\vec B})}(\hat {\vec A})$, ${\vec B}'(t,{\vec Q},{\vec P})=W_{(t,P_t,{\vec Q},{\vec P})}^{(t,\phi,{\vec A},{\vec B})}(\hat {\vec B})$ and $\phi'(t,P_t,{\vec Q},{\vec P})=W_{(t,P_t,{\vec Q},{\vec P})}^{(t,\phi,{\vec A},{\vec B})}(\hat{\phi} )$. Finally, let us concisely designate the transformation (41) by $\vec T: \R^{(2N+2)} \longrightarrow \R^{(2N+2)};\; (t,P_t,{\vec Q},{\vec P}) \longrightarrow (t,\phi,{\vec A},{\vec B})=\vec T(t,P_t,{\vec Q},{\vec P})$.

The transformation $\vec T$ is canonical but in general it does not preserve neither the starproduct nor the Moyal bracket. 
Using the generalized Weyl transform $W_{(t,P_t,{\vec Q},{\vec P})}^{(t,\phi,{\vec A},{\vec B})}$ we find a new phase space representation of the eigenvalue equation (18): 
\begin{equation}
\phi'(t,P_t,{\vec Q},{\vec P}) *'_{(t,P_t,{\vec Q},{\vec P})} f'_W (t,P_t,{\vec Q},{\vec P}) = f'_W(t,P_t,{\vec Q},{\vec P}) *'_{(t,P_t,{\vec Q},{\vec P})} \phi'(t,P_t,{\vec Q},{\vec P}) =0
\end{equation}
where the $*'_{(t,P_t,{\vec Q},{\vec P})}$ is the covariant starproduct given by (3) with:
\begin{equation}
J'^{ij}_{(t,P_t,{\vec Q},{\vec P})}=J^{ij}_{(t,\phi ,{\vec A},{\vec B})} \quad \mbox{and} \quad \Gamma'^i_{jk} =  \frac{\partial O'^i}{\partial O^b} \frac{\partial^2 O^b}{\partial O'^j \partial O'^k},
\end{equation}
where $O^b \in \{t,\phi ,{\vec A},{\vec B}\}$ and $O'^i \in \{t, P_t,{\vec Q},{\vec P}\}$.
The solutions of eq.(43) can be read from eq.(23,28):
\begin{equation}
f'_W(t,P_t,{\vec Q},{\vec P}) =\frac{1}{(2\pi \hbar)^N} W_{(t,P_t,{\vec Q},{\vec P})}^{(t,\phi,{\vec A},{\vec B})}(|\psi><\psi|)= \frac{1}{(2\pi \hbar)^N} \int d{\vec a} \int d{\vec b} C({\vec a}) C^* ({\vec b}) \rho'_{{\vec a},{\vec b}} (t,P_t,{\vec Q},{\vec P}),
\end{equation}
where:
\begin{eqnarray}
&&\rho'_{{\vec a},{\vec b}}(t,P_t,{\vec Q},{\vec P}) = \Delta_{*'_{(t,P_t,{\vec Q},{\vec P})}}(\phi'(t,P_t,{\vec Q},{\vec P}),{\vec A}'(t,{\vec Q},{\vec P});h=0,{\vec a};h=0,{\vec b})  \nonumber \\
&=& \Delta_{*'_{(t,P_t,{\vec Q},{\vec P})}}(\phi'(t,P_t,{\vec Q},{\vec P})) {\prod_{j=1}^N}{}_{*'_{(t,P_t,{\vec Q},{\vec P})}} \Delta_{*'_{(t,P_t,{\vec Q},{\vec P})}}(A_j'(t,{\vec Q},{\vec P}),a_j,b_j) =\nonumber \\
& = & \Delta _{*_{(t,\phi,{\vec A},{\vec B})}}(\phi'(t,P_t,{\vec Q},{\vec P}) ){\prod_{j=1}^N}{}_{*_{(t,\phi,{\vec A},{\vec B})}}\left\{ e_{*_{(t,\phi,{\vec A},{\vec B})}}^{\frac{i}{\hbar} (b_j-a_j)
B'_j(t,{\vec Q},{\vec P})} *_{(t,\phi,{\vec A},{\vec B})} \Delta _{*_{(t,\phi,{\vec A},{\vec B})}}(A'_j(t,{\vec Q},{\vec P})-b_j) \right\}\nonumber \\
& = & \delta (\phi' (t,P_t,{\vec Q},{\vec P}) )\prod_{j=1}^N \left\{e^{\frac{i}{\hbar} (b_j-a_j)B'_j(t,{\vec Q},{\vec P})}  \delta (A_j'(t,{\vec Q},{\vec P})-\frac{a_j+b_j}{2})\right\} \nonumber \\
& = & \delta (\phi' (t,P_t,{\vec Q},{\vec P}) ) e^{\frac{i}{\hbar} (\vec b-\vec a)\cdot \vec B'(t,{\vec Q},{\vec P})}  \delta (\vec A'(t,{\vec Q},{\vec P})-\frac{\vec a+\vec b}{2})  
\end{eqnarray}
The new Wigner function satisfies:
\begin{equation}
f_W'(t,P_t,{\vec Q},{\vec P})=f_W(t,\phi '(t,P_t,{\vec Q},{\vec P}),\vec A'(t,{\vec Q},{\vec P}),\vec B'(t,{\vec Q},{\vec P}))
\end{equation}
where $f_W$ is given by eq.(28). 
Furthermore and since the coordinate transformation is canonical we have:
\begin{equation}
d\mu=d\phi d{\vec A}d{\vec B}=  \delta (t) dt  d\phi d{\vec A}d{\vec B}= \delta (t) dt dP_td{\vec Q}d{\vec P}=dP_td{\vec Q}d{\vec P},
\end{equation}
and the new Wigner function obeys the proper normalization: $\int d \mu f'_W = 1$.
Also notice that both the diagonal and the non-diagonal stargenfunctions (and thus also the Wigner function) have support only on the classical constraint hypersurface a property that it is not shared by the previous representation (36,37).

Finally, let us briefly elaborate on the dynamical structure of the system in this representation. 
It is clear from eq.(43) that the Wigner function is static:
\begin{eqnarray}
\frac{\partial}{\partial \tau} f'_W(t,P_t,{\vec Q},{\vec P};\tau) & = & [H'(t,P_t,{\vec Q},{\vec P}),f'_W(t,P_t,{\vec Q},{\vec P};\tau)]_{M'_{(t,P_t,{\vec Q},{\vec P})}} \nonumber \\
& = & \lambda [\phi' (t,P_t,{\vec Q},{\vec P}),f'_W(t,P_t,{\vec Q},{\vec P};\tau)]_{M'_{(t,P_t,{\vec Q},{\vec P})}} =0,
\end{eqnarray}
confirming the typical picture of frozen dynamics. However, the Wigner function does evolve with respect to the physical time $t$. From eqs.(42,47) we have:
\begin{eqnarray}
\frac{\partial}{\partial t} f'_W(t,P_t,{\vec Q},{\vec P}) & = & \frac{\partial f_W}{\partial \phi}(\vec T(t,P_t,{\vec Q},{\vec P}))\frac{\partial \phi'}{\partial t}+\sum_{j=1}^N \frac{\partial f_W}{\partial A_j}(\vec T(t,P_t,{\vec Q},{\vec P}))\frac{\partial A'_j}{\partial t} + \sum_{j=1}^N \frac{\partial f_W}{\partial B_j}(\vec T(t,P_t,{\vec Q},{\vec P}))\frac{\partial B'_j}{\partial t}  \nonumber \\
&=& \sum_{j=1}^N \frac{\partial f_W}{\partial A_j} (\vec T(t,P_t,{\vec Q},{\vec P}))\{H_0,A'_j\}_{(t,P_t,{\vec Q},{\vec P})} + \sum_{j=1}^N \frac{\partial f_W}{\partial B_j}(\vec T(t,P_t,{\vec Q},{\vec P})) \{H_0,B'_j\}_{(t,P_t,{\vec Q},{\vec P})} \nonumber \\
&=& \{H_0,f'_W\}_{(t,P_t,{\vec Q},{\vec P})},  
\end{eqnarray}
reproducing the classical Liouville equation $\{H',f'_W\}_{(t,P_t,{\vec Q},{\vec P})} =0$. 

We see that the $^{(t,\phi , {\vec A},{\vec B})}_{(t,P_t,{\vec Q},{\vec P})}$-representation of the system leads to an interesting mathematical picture: the Hamiltonian vector field lives on the extended phase space and is given by:
\begin{equation}
\xi_H = \lambda \left\{ \frac{\partial}{\partial t} + \sum_{j=1}^N \frac{\partial H_0}{\partial P_j} \frac{\partial}{\partial Q_j} -\sum_{j=1}^N 
\frac{\partial H_0}{\partial Q_j} \frac{\partial}{\partial P_j} \right\}.
\end{equation}
The flows of this vector field define lines in phase space. These lines cross each time hypersurface once and only once: they are the histories of the system. These histories are one-dimensional hypersurfaces that can be identified by the values of the $1+2N$ constants of motion: $\phi, {\vec A}, {\vec B}$. Along these lines the Wigner distribution function is constant. However, the correlations between the canonical variables, namely between ${\vec Q}$ and $t$, do change and time evolution is imprint on these correlations. The interesting point is that this picture is not of the classical description but of the quantum mechanical instead. 

A natural question is then: what happened to the quantum content of the theory? What happened to the interfering trajectories and to the non-local behavior? The answer is that the quantum features have been completely removed from the distributional sector and are now exclusively placed in the observables' sector of the theory. This can be checked explicitly by applying the proper generalized Weyl transform to one of the fundamental operators of the system. We have for instance:
\begin{equation}
W_{(t,P_t,{\vec Q},{\vec P})}^{(t,\phi,{\vec A},{\vec B})} (\hat q_1)=q_1(t,\phi,{\vec A},{\vec B})|_{(t,\phi,{\vec A},{\vec B})=\vec T(t,P_t,{\vec Q},{\vec P})}=q_1(t,\phi' (t,P_t,{\vec Q},{\vec P}),A'(t,{\vec Q},{\vec P}),B'(t,{\vec Q},{\vec P})),
\end{equation}
leading to the time evolution:
\begin{eqnarray}
\frac{dq_1}{dt}& = & \frac{\partial q_1}{\partial t}+ \frac{\partial q_1}{\partial \phi}\frac{\partial \phi'}{\partial t} +\sum_{j=1}^N
\frac{\partial q_1}{\partial A_j}\frac{\partial A'_j}{\partial t} + \sum_{j=1}^N\frac{\partial q_1}{\partial B_j} \frac{\partial B'_j}{\partial t} \nonumber \\
& = & [q_1,H_0]_{M_{({\vec A},{\vec B})}}-\sum_{j=1}^N\frac{\partial q_1}{\partial A_j} \{A'_j,H_0 \}_{({\vec Q},{\vec P})} -\sum_{j=1}^N \frac{\partial q_1}{\partial B_j}
\{B'_j,H_0 \}_{({\vec Q},{\vec P})} \nonumber \\
&=&   [q_1,H_0]_{M_{({\vec A},{\vec B})}}-\{q_1,H_0 \}_{({\vec A},{\vec B})}, 
\end{eqnarray}
where the identity $\frac{\partial q_1}{\partial t}= [q_1,H_0]_{M_{({\vec A},{\vec B})}}$ follows from eq.(20) (or alternatively from eq.(35)), we used the fact that $\phi '$ is time independent and that the transformation $({\vec Q},{\vec P}) \to ({\vec A},{\vec B})$ is canonical.  Equation (53) does in fact constitute a quantum correction to the classical statistical description where we have:
\begin{equation}
\frac{\partial \rho}{\partial t}= \{H_0,\rho \} \quad \mbox{and} \quad \frac{d q_1}{d t}=0 .
\end{equation}
Notice that in the classical description ${\vec Q}={\vec q}$,  ${\vec P}={\vec p}$ and ${\vec q}(t,\phi' (t,P_t,{\vec q},{\vec p}),{\vec A}'(t,{\vec q},{\vec p}),{\vec B}'(t,{\vec q},{\vec p}))={\vec q}$.

We finally conclude by pointing out that a causal representation can also be obtained for deparametrized systems by using an explicit "time dependent phase space representation" where the starproduct and Moyal bracket are themselves (scalar) time dependent \cite{Nuno7}. In this approach we apply a time dependent generalization of the Weyl map to the density matrix formulation of the deparametrized system and obtain a phase space causal representation of the Wigner function, which is formally identical to the one described by eqs.(45,46,52), although $t$ is a canonical variable in the former equations and an external scalar parameter in the approach of \cite{Nuno7}. Accordingly, the two resulting quasidistributions live on different phase spaces.
In spite of this the two formulations are consistent with each other leading to the single conclusion that a phase space formulation of quantum mechanics where the distributional sector displays a fully classical causal structure is made possible through a suitable choice of representation. This may either be a (scalar) time dependent representation in the deparametrized phase space or a history representation in the generally covariant setting.

\section{Example}

To illustrate our previous results let us consider the simple system compose by two coupled particles and described by the Hamiltonian:
\begin{equation}
H_0= \frac{ p_1^2}{2M} + \frac{ p_2^2}{2m} + k  q_1 p_2^2
\end{equation}
where $( q_1, p_1)$ are the canonical variables of the particle of mass $M$, $(q_2,p_2)$ those of the particle of mass $m$ and $k$ is a coupling constant. 

The generally covariant version of this system is obtained by promoting $t$ to a canonical variable and imposing the time reparametrization invariance. The 
extended Hamiltonian formulation of the system lives on a $6$-dimensional phase space spanned by the canonical variables $t,P_t,\vec q=(q_1,q_2),\vec p=(p_1,p_2)$ which satisfy the commutation relations $\{t,P_t\}=1$, $\{q_i,p_j\}=\delta_{ij}$, $i,j=1,2$, all others being zero. Upon quantization the
system is described by the "zero Hamiltonian":
\begin{equation}
\hat H=\lambda \hat{\phi} \quad , \quad \hat{\phi}=\hat{P}_t+\hat H_0
=\hat{P}_t+\frac{\hat p_1^2}{2M} + \frac{\hat p_2^2}{2m} + k  \hat q_1\hat p_2^2
\end{equation}
where $\hat{\phi}$ is the first class Hamiltonian constraint and $\lambda$ is a Lagrange multiplier. In standard the Dirac formulation the physical states of this system are the wave functions $\psi$ solutions of the constraint equation:
\begin{equation}
\hat{\phi} |\psi>=0
\end{equation}

We now address the deformation quantization of the system. As in the main text, three distinct quantum phase space representations will be presented.
\\

{\bf 1.} {\it The map $W_{(t,P_t,\vec q,\vec p)}$ and the generally covariant Schr\"odinger picture}.

The Weyl map $W_{(t,P_t,\vec q,\vec p)}$ yields the $(t,P_t,\vec q,\vec p)$-constraint symbol:
\begin{equation}
\phi= W_{(t,P_t,\vec q,\vec p)}(\hat{\phi})=P_t+\frac{ p_1^2}{2M} + \frac{ p_2^2}{2m} + k q_1 p_2^2
\end{equation}
and the quantum phase space version of the Hamiltonian constraint (57) is given by:
\begin{equation}  
\phi (t,P_t,\vec q,\vec p) *_{(t,P_t,\vec q,\vec p)}f_W(t,P_t,\vec q,\vec p) =f_W(t,P_t,\vec q,\vec p) *_{(t,P_t,\vec q,\vec p)} \phi (t,P_t,\vec q,\vec p)=0
\end{equation}
The solutions of this right and left stargenvalue equation are given by eqs.(36,37)
and they do not easily simplify any further. It is worth noticing that due to the non-local character of the $*$-product the phase space quasidistribution $f_W$ will have support on phase-space regions which are not classically allowed. One can easily calculate the evolution of $f_W$ with respect to the physical time $t$. From eq.(59) we have:
\begin{eqnarray}
&&\frac{\partial f_W}{\partial \tau}=\lambda [\phi,f_W]_{M_{(t,P_t,\vec q,\vec p)}}=0 \Longleftrightarrow 
\frac{\partial f_W}{\partial t}=[H_0,f_W]_{M_{(\vec q,\vec p)}}\nonumber \\ &\Longleftrightarrow &\frac{\partial f_W}{\partial t}=\{H_0,f_W\}_{{(\vec q,\vec p)}}
+\frac{\hbar^2}{24} \left[2\left\{2kp_2,\frac{\partial^2 f_W}{\partial q_2 \partial p_1}\right\}_{(\vec q,\vec p)} -
\left\{2kq_1,\frac{\partial^2 f_W}{\partial q_2^2}\right\}_{(\vec q,\vec p)} \right]
\end{eqnarray}
which is obviously not of the form of the Liouville equation. Consequently, the Wigner function does not display a classical causal structure.\\

{\bf 2.} {\it The map $W_{(t,\phi,\vec A,\vec B)}$ and the generally covariant Heisenberg picture}.

Following the approach of section 3 our first step is to determine the quantum histories associated to $\hat H$. We easily find that [check eqs.(19,20)]:
\begin{equation}
\left\{ \begin{array}{lll}
\hat A_1 & = & \hat q_1 - \frac{\hat p_1}{M} \hat t - \frac{k}{2M} \hat p_2^2\hat t^2 \\
\hat B_1 & = & \hat p_1+k\hat p_2^2\hat t \\
\hat A_2 & = & \hat q_2 - \left\{\frac{\hat p_2}{m} + 2k\hat q_1\hat p_2 \right\}\hat t + \frac{k}{M} \hat p_1\hat p_2\hat t^2 + \frac{k^2}{3M}\hat p_2^3\hat t^3 \\
\hat B_2 & = &\hat p_2
\end{array} \right.
\end{equation}
satisfy the requisites following eq.(18) i.e. $\hat A_j$, $\hat B_j$, $j=1,2$ commute with the constraint $\hat{\phi}$ and with $\hat t$ and furthermore they satisfy the Heisenberg algebra $[\hat A_1,\hat B_1]=[\hat A_2,\hat B_2]=i\hbar$, all other commutators being zero. Since $[\hat t,\hat{\phi}]=i\hbar$ the set 
$\{\hat t,\hat{\phi},\hat{\vec A}=(\hat A_1,\hat A_2),\hat{\vec B}=(\hat B_1,\hat B_2)\}$ is a complete set of fundamental operators for this system. 

Using the Weyl map $W_{(t,\phi,\vec A,\vec B)}$ we find the Heisenberg phase space representation of the constraint equation (57):   
\begin{equation}
\phi *_{(t,\phi,\vec A,\vec B)} f_W(t,\phi,\vec A,\vec B)=f_W(t,\phi,\vec A,\vec B)*_{(t,\phi,\vec A,\vec B)}\phi=0
\end{equation}
The fundamental solutions of this equation $\rho_{\vec a,\vec b}$ (where $\vec a=(a_1,a_2)$ and $\vec b=(b_1,b_2)$), and also satisfying:
\begin{equation}
A_j *_{(t,\phi,\vec A,\vec B)} \rho_{\vec a,\vec b}= a_j \rho_{\vec a,\vec b} \quad , \quad 
\rho_{\vec a,\vec b}*_{(t,\phi,\vec A,\vec B)} A_j=b_j \rho_{\vec a,\vec b}
\quad, \quad j=1,2
\end{equation}
are given by:
\begin{equation}
\rho_{\vec a,\vec b}(t,\phi,\vec A,\vec B)=\delta(\phi) e^{\frac{i}{\hbar}\{ (b_1- a_1) B_1+(b_2-a_2)B_2 \}} \delta( A_1-\frac{ a_1+ b_1}{2})\delta( A_2-\frac{ a_2+ b_2}{2})
\end{equation}
And the Wigner function, solution of (62), is just a hermitian combination of the fundamental solutions. We have:
\begin{equation}
f_W(t,\phi,\vec A,\vec B) =\frac{1}{2\pi \hbar}  \int d\vec a \int d\vec b C(\vec a) C^* (\vec b) \rho_{\vec a,\vec b} (t,\phi,\vec A,\vec B)
\end{equation}
where $C(\vec a)$ obeys to the normalization condition that is induced by the normalization of the Wigner function: $\int d\vec a |C(\vec a)|^2=1$.
The Wigner function satisfies eq.(62) which implies that $\frac{\partial f_W}{\partial \tau}=0$. We also have from eq.(64) that $\frac{\partial f_W}{\partial t}=0$. That is, the Wigner function is static both with respect to the unphysical scalar parameter as well as to the physical time. This is an expected result since, in this representation, the Wigner quasidistribution is an exclusive function of observables of the system (the histories) i.e. of quantities that commute with the Hamiltonian constraint.

On the other hand, in this representation, the stargenfunctions of the fundamental variables $\hat t,\hat{\vec q}$ and $\hat{\vec p}$ do evolve with respect to the physical time. For instance (let $|x>$ be the general eigenket of $\hat q_1$ with associated eigenvalue $x$):
\begin{eqnarray}
g_{x}(t,\phi,\vec A,\vec B) & = & W_{(t,\phi,\vec A,\vec B)} (|x><x|)  =  \Delta_{*_{(t,\phi,\vec A,\vec B)}} (q_1(t,\vec A,\vec B)-x)  \\
&=& \frac{1}{2\pi} \int d k \, e_{*_{(t,\phi,\vec A,\vec B)}}^{ik(q_1(t,\vec A,\vec B)-x)}=\frac{1}{2\pi} \int d k \, e^{ik(q_1(t,\vec A,\vec B)-x)}=\delta (q_1(t,\vec A,\vec B)-x) \nonumber
\end{eqnarray}
where $q_1(t,\vec A,\vec B)$ can be easily derived from eq.(61): $q_1(t,\vec A,\vec B)=W_{(t,\phi,\vec A,\vec B)}(\hat q_1(\hat t,\hat{\vec A},\hat{\vec B}))= A_1 + \frac{ B_1}{M} t - \frac{k}{2M}  B_2^2 t^2$. Hence, $g_{x}(t,\phi ,\vec A,\vec B)$ satisfies:
\begin{equation}
\frac{\partial }{\partial t} g_{x}(t,\phi ,\vec A,\vec B) =  [g_{x}(t,\phi ,\vec A,\vec B),H_0(\vec A,\vec B)]_{M_{(\vec A,\vec B)}}=
\{g_{x}(t,\phi ,\vec A,\vec B),H_0(\vec A,\vec B)\}_{(\vec A,\vec B)},
\end{equation}
where $H_0(\vec A,\vec B)=W_{(\vec A,\vec B)}(\hat H_0)=
 \frac{ B_1^2}{2M} + \frac{ B_2^2}{2m} + k  A_1 B_2^2$.
We conclude that the Weyl transform $W_{(t,\phi,\vec A,\vec B)}$ casts the phase space dynamics in the Heisenberg picture. Accordingly, the time dependence is exclusively displayed by the observable (stargenfunction) sector of the theory.\\

{\bf 3.} {\it The map $W_{(t,P_t,\vec q,\vec p)}^{(t,\phi,\vec A,\vec B)}$ and the generally covariant causal picture}.

Following the prescription of section 4 let us introduce a new set of  phase space coordinates $(t,P_t,\vec Q=(Q_1,Q_2),\vec P=(P_1,P_2))$ defined by eqs.(39,40). We get: $t=t$, $P_t=\phi-H_0(\vec A,\vec B)$ and,
\begin{equation}
\left\{ \begin{array}{lll}
Q_1 & = & A_1 + \frac{B_1}{M} t - \frac{k}{2M}  B_2^2 t^2 \\
 P_1 & = &  B_1-k B_2^2 t \\
 Q_2 & = &  A_2 + \left\{\frac{ B_2}{m} + 2k A_1 B_2 \right\} t \\
&& + \frac{k}{M}  B_1 B_2 t^2 - \frac{k^2}{3M} B_2^3 t^3 \\
 P_2 & = & B_2
\end{array} \right.
\Longleftrightarrow
\left\{ \begin{array}{lll}
A_1 & = & Q_1 - \frac{P_1}{M} t - \frac{k}{2M}  P_2^2 t^2 \\
 B_1 & = &  P_1+k P_2^2 t \\
 A_2 & = &  Q_2 - \left\{\frac{ P_2}{m} + 2k Q_1 P_2 \right\} t \\
&& + \frac{k}{M}  P_1 P_2 t^2 + \frac{k^2}{3M} P_2^3 t^3 \\
 B_2 & = & P_2
\end{array} \right.
\end{equation}
We notice that $\vec Q(t,\vec A,\vec B)$ and $\vec P(t,\vec A,\vec B)$ coincide with $W_{(t,\phi,\vec A,\vec B)}(\hat{\vec q})$, $W_{(t,\phi,\vec A,\vec B)}(\hat{\vec p})$, respectively (this is an easy result that follows from eq.(61)), i.e the classical and the quantum histories of this system are exactly the same. Indeed, eq.(68) solves both the Moyal and the Hamiltonian equations of motion. In the notation of section 4 we have $\vec A'=\vec A$ and  $\vec B'=\vec B$ (one should notice that this result is not valid in general). Hence, for this system, we are not required to introduce a second set of "classical coordinates" and can simplify the notation by making:  $Q_j=q_j$ and $P_j=p_j$, $j=1,2$. 

We now consider 
the action of the generalized Weyl map $W_{(t,P_t,\vec q,\vec p)}^{(t,\phi,\vec A,\vec B)}$. The associated covariant starproduct $*'_{(t,P_t,\vec q,\vec p)}$ and Moyal bracket $[\, , \, ]_{M_{(t,P_t,\vec q,\vec p)}}$ are characterized by (let $O'^1=P_t,\,O'^2=p_1,\, O'^3=p_2,\,O'^4=t,\, O'^5=q_1,\,O'^6=q_2$, 
$O^1=\phi ,\,O^2=B_1,\, O^3=B_2,\,O^4=t,\, O^5=A_1,\,O^6=A_2$ and $i,j=1..6$) [check eq.(5)]:
\begin{eqnarray}
& J'^{ij}_{(\vec q,\vec p)} =  J^{ij}_{(\vec q,\vec p)} & \\
& \Gamma'^1_{22}=-\Gamma'^5_{24}=-\Gamma'^5_{42}=\frac{1}{M}, \quad \Gamma'^1_{24}=\Gamma'^1_{42}=-\Gamma'^5_{44}=\frac{k}{M}p_2^2, \quad \Gamma'^1_{33}=2kA_1(t,\vec q,\vec p), & \nonumber \\
& \Gamma'^1_{34}= \Gamma'^1_{43}=-\Gamma'^6_{44}=-\frac{2k}{M}p_1p_2, 
\quad \Gamma'^1_{35}=\Gamma'^1_{53}=\Gamma'^2_{34}=\Gamma'^2_{43}=-\Gamma'^6_{45}=-\Gamma'^6_{54}=2kp_2,& \nonumber \\
& \Gamma'^1_{44}= \frac{k^2}{M}p_2^4, \quad \Gamma'^2_{33}=-\Gamma'^6_{35}=-\Gamma'^6_{53}= 2kt, \quad 
 \Gamma'^5_{33}=\Gamma'^6_{23}=\Gamma'^6_{32}=\frac{k}{M}t^2 ,& \nonumber \\
& \Gamma'^5_{34}=\Gamma'^5_{43}=\frac{k}{M}p_2t ,\quad 
\Gamma'^6_{33}=\frac{2k^2}{M}p_2t^3 ,\quad \Gamma'^6_{34}=\Gamma'^6_{43}=-\frac{1}{m}-2kA_1(t,\vec q,\vec p) & \nonumber
\end{eqnarray}
all other Christoffel symbols being zero. 

In this representation the constraint equation reads:
\begin{equation}
(P_t+H_0(\vec q,\vec p)) *'_{(t,P_t,\vec q,\vec p)} f'_W (t,P_t,\vec q,\vec p) = f'_W(t,P_t,\vec q,\vec p) *'_{(t,P_t,\vec q,\vec p)}  (P_t+H_0(\vec q,\vec p))=0
\end{equation}
where $H_0(\vec q,\vec p)=W_{(t,P_t,\vec q,\vec p)}^{(t,\phi,\vec A,\vec B)}(\hat H_0)$ is given by eq.(54). The fundamental solutions of eq.(70) are given by:
\begin{eqnarray}
&&\rho'_{\vec a,\vec b}(t,P_t,\vec q,\vec p) = \Delta_{*'_{(t,P_t,\vec q,\vec p)}}(\phi(t,P_t,\vec q,\vec p),\vec A(t,\vec q,\vec p);h=0,\vec a;h=0,\vec b)  \nonumber \\
& = & \delta (\phi (t,P_t,\vec q,\vec p) ) e^{\frac{i}{\hbar} (\vec b-\vec a)\cdot\vec B(t,\vec q,\vec p)}  \delta (\vec A(t,\vec q,\vec p)-\frac{\vec a+\vec b}{2}).  \nonumber \\
&=& \delta (P_t+H_0(\vec q,\vec p) ) e^{\frac{i}{\hbar} \{(b_1- a_1) (p_1+kp_2^2t) + (b_2-a_2) p_2 \}}  \delta \left(q_1 - \frac{p_1}{M} t - \frac{k}{2M}  p_2^2 t^2 -\frac{a_1+b_1}{2}\right)\nonumber \\
&& \cdot \delta \left( q_2 - \left\{\frac{ p_2}{m} + 2k q_1 p_2 \right\} t 
 + \frac{k}{M}  p_1 p_2 t^2 + \frac{k^2}{3M} p_2^3 t^3 
-\frac{a_2+b_2}{2}\right)
\end{eqnarray}
and the Wigner function is once again a hermitian combination of the fundamental solutions $\rho'_{\vec a,\vec b}$: 
\begin{equation}
f'_W(t,P_t,\vec q,\vec p) =\frac{1}{(2\pi \hbar)^N} W_{(t,P_t,\vec q,\vec p)}^{(t,\phi,\vec A,\vec B)}(|\psi><\psi|)= \frac{1}{(2\pi \hbar)^N} \int d\vec a \int d\vec b C(\vec a) C^* (\vec b) \rho'_{\vec a,\vec b} (t,P_t,\vec q,\vec p),
\end{equation}
It is related with the $(t,\phi,\vec A,\vec B)$-representation by:
\begin{equation}
f'_W(t,P_t,\vec q,\vec p)=f_W(0,\phi(t,P_t,\vec q,\vec p),\vec A(t,\vec q,\vec p),\vec B(t,\vec q,\vec p))
\end{equation}
where $f_W$ is given by eqs.(64,65). We conclude that: 1) The support of $f_W'$ is confined to the classically allowed regions [check eq.(71)] and 2) its evolution with respect to the physical time satisfies
the Liouville equation:
\begin{equation}
\frac{\partial f'_W}{\partial t}=\sum_{j=1}^2 \frac{\partial f_W}{\partial A_j}  \frac{\partial A_j}{\partial t} +\sum_{j=1}^2
\frac{\partial f_W}{\partial B_j}  \frac{\partial  B_j}{\partial t}
=\sum_{j=1}^2 \frac{\partial f_W}{\partial A_j} \{H_0,A_j\}_{(\vec q,\vec p)}+\sum_{j=1}^2\frac{\partial f_W}{\partial B_j} \{H_0,B_j\}_{(\vec q,\vec p)} =\{H_0,f'_W\}_{(\vec q,\vec p)}
\end{equation}
Hence, in this representation the quantum behavior is displayed by the stargenfunction sector alone.
However, for this system, we also have (let $z=q_1,p_1 \vee p_2$ and $|z_0>$ be a generic eigenket of $\hat z$ with associated eigenvalue $z_0$):
\begin{eqnarray}
W_{(t,P_t,\vec q,\vec p)}^{(t,\phi,\vec A,\vec B)}(|z_0><z_0|) & = & \Delta_{*'(t,P_t,\vec q,\vec p)}(z-z_0)=\Delta_{*(t,\phi,\vec A,\vec B)}(z(t,\vec A,\vec B)-z_0)|_{\vec A=\vec A(t,\vec q,\vec p) \wedge \vec B=\vec B(t,\vec q,\vec p)} \nonumber \\
& = & \delta(z(t,\vec A,\vec B)-z_0)|_{\vec A=\vec A(t,\vec q,\vec p) \wedge \vec B=\vec B(t,\vec q,\vec p)}= \delta(z-z_0)
\end{eqnarray}
where in the third step we used the fact that  $e_{*_{(t,\phi,A,B)}}^{ik(z(t,\vec A,\vec B)-z_0)}= e^{ik(z(t,\vec A,\vec B)-z_0)}$ for $z=q_1,p_1 \vee p_2$, (a simple result that follows from eq.(61)).
Hence, the former three fundamental stargenfunctions display a classical structure and satisfy $\frac{\partial}{\partial t}\Delta_{*'(t,P_t,q,p)}(z-z_0)=0$. We conclude that for this system, in this representation, the non-trivial (quantum) behavior is displayed by the stargenfunction $z=q_2$ alone.

\section{Conclusions}

We addressed the deformation quantization of the non-relativistic particle in the generally parametrized form and provided the complete specification of its physical quantum states in three different phase space representations. 
We proved that in one of these representations the distributional sector of the system displays a classical causal structure. This result confirms a similar conclusion that has been recently obtained for a generic non-relativistic deparametrized system and reinforces the point of view that the De Broglie-Bohm formulation is not the unique possible causal formulation of quantum mechanics. 
In the De Broglie-Bohm theory \cite{Holland,Nuno8,Nuno9} the source of quantum behavior is the quantum potential determining a causal (although not fully classical) dynamics for the quasidistribution. Furthermore, the theory also displays a non-trivial quantum correction to the momentum stargenfunction. On the other hand, in the "causal covariant formulation" presented here the quantum effects have been completely removed from the distributional sector (which now displays a fully classical causal structure) and the price to pay was the appearance of some further (quantum) corrections on the observable's sector of the theory. 

We finish by recalling the point of view according to which the relation between the classical statistical and the quantum mechanical formulations of generally covariant systems should be further explored as it may provide key physical insights into some of the conceptual problems displayed by the quantum version of these systems. 
From this perspective the deformation methods seem to be especially suited. We proved that (at least for the simplest case of the parametrized non-relativistic particle) the formal similarities between classical statistical and phase space quantum mechanics carry on intact to the generally covariant context. This close analogy is obviously superior to the one displayed by the standard operator formulation, and supports the point of view that the deformation methods should be further explored as an alternative, conceptually simpler approach to the quantization of generally covariant systems.

\paragraph*{Acknowledgments.} 

 This work was partially supported by the grant POCTI/MAT/45306/2002.

\end{document}